\documentclass[fleqn,usenatbib,useAMS]{mnras}

\usepackage{graphicx}	
\usepackage{amsmath}	
\usepackage{amssymb}	
\usepackage{multicol}      	
\usepackage{bm}		
\usepackage{pdflscape}	
\usepackage{float}
\usepackage{xcolor}

\usepackage[T1]{fontenc}
\usepackage{ae,aecompl}

\usepackage{newtxtext,newtxmath}

\title[Spectra of cool  DQ and DQp white dwarfs]{Consistent multi-wavelength spectral modeling of cool white dwarfs with carbon-polluted atmospheres}
\author[Farihi et al.]{Jay Farihi$^1$\thanks{E-mail:  j.farihi@ucl.ac.uk},
Piotr M.~Kowalski$^2$\thanks{E-mail:  p.kowalski@fz-juelich.de},  
Sandy K.~Leggett$^3$,
Hania Azzam$^2$,
Jackson Headon$^{2,4}$, and John P.~
\newauthor
Subasavage$^5$
\\
$^1$Department of Physics and Astronomy, University College London, London WC1E 6BT, UK\\
$^2$Institute of Energy Technologies: Theory and Computation of Energy Materials (IET	-3), Forschungszentrum J\"ulich GmbH, 52425 J\"ulich, Germany\\
$^3$Gemini Observatory, 670 N. A\'ohoku Place, Hilo HI 96720, USA\\
$^4$Department of Physics, Arizona State University, Tempe, AZ 85281, USA\\
$^5$The Aerospace Corporation, 2310 E. El Segundo Boulevard, El Segundo, CA 90245, USA}

\begin{document}


\maketitle

\begin{abstract}
This work presents atmospheric modeling of multi-wavelength spectra for eight cool white dwarfs with carbon-enriched atmospheres, with four each of spectral type DQ and DQp.  The Deslanders-d'Azambuja bands of C$_2$ are detected in all six DQ stars that have data covering wavelengths shorter than 4000\,\AA.  These bands are blue shifted in the DQp types, consistent with that observed for the Swan bands, and reproduced with the same pressure distortion model. The coolest stars in the sample show significant near-infrared flux suppression, identified here as collision-induced absorption from dense helium, with trace hydrogen [H/He] $<-6$, estimated from the absence of CH features.  Notably, the near-ultraviolet through near-infrared spectral energy distribution of WD\,0038$-$226 is correctly reproduced using an atmospheric model with [H/He] $=-6.8$; however, beyond 2\,$\upmu$m the predicted flux is too high, and a carbon-free atmosphere is able to reproduce the mid-infrared observations well.  The multi-wavelength coverage permits accurate modeling characterization of these stars, including constraining the ionization equilibrium in fluid helium and the resulting density profiles. The densities found in this work are 2--3 times higher than determined in previous studies, approaching 2\,g\,cm$^{-3}$ for the coolest stars. Excluding the uncertain high masses obtained for the two coolest DQp stars, the sample stars have masses that are broadly consistent with those found for the wider white dwarf population, with no difference across the DQ-DQp temperature boundary. 

\end{abstract}

\begin{keywords}		
    stars: abundances--- 
    stars: atmospheres--- 
    stars: chemically peculiar---
    stars: evolution---
    white dwarfs
\end{keywords}

\section{Introduction}

Cool white dwarfs represent the remnants of some of the oldest stars in the Galaxy.  It has been shown that complementary studies using observations and atmospheric modeling for large samples of white dwarfs can reveal vital information on the formation and evolution of stars \citep{campos2016,cheng2020,bedard2024a}, their planetary systems \citep{zuckerman2003,farihi2009a,doyle2023}, and provide an accurate tool for measuring the ages of stellar populations through white dwarf cosmochronology \citep[e.g.][]{fontaine2001,garcia-berro2010,kilic2019a,moss2022}.  

Among $T_{\rm eff} \lesssim 7000$\,K white dwarfs that generally represent stars with total ages significantly greater than 1\,Gyr, the DQ and DQp (p= peculiar) spectral classes exhibit normal or distorted C$_2$ Swan bands in their optical spectra, respectively.  In contrast to featureless DC white dwarfs, these spectral features provide an empirical handle for atmospheric models, which help to constrain stellar and spectral evolution, and connect observables to atmospheric and interior stellar models. The classical, cool DQ stars have been successfully modeled in myriad studies with helium-dominated atmospheres enriched with trace carbon \citep[e.g.][]{dufour2005,koester2006b,blouin2019b}, whose presence is the result of core dredge-up by the deep helium convection zone \citep{pelletier1986,koester2020,blouin2023a,blouin2023b}.  There are two interesting phenomena observed in the local samples of DQ stars: (1) their cooling sequence appears to abruptly end near $T_{\rm eff}\approx 6000$\,K \citep{dufour2005,koester2019} and (2) the DQp stars appear at lower effective temperature, where their cooling sequence is an apparent continuation of DQ star evolution \citep{kowalski2013,blouin2019b}.  The known DQp stars typically have Swan band minima that are blue-shifted by $\approx$150\,\AA\ \citep{schmidt1995,bergeron1997,hall2008a}.

\begin{table*}
\caption{Spectroscopic observation log.
\label{obs}}
\begin{tabular}{@{}lllcllccccccc@{}}

\hline

WD			&Other Name 	&SpT	&$G$ 	&Instrument	&Date	&\multicolumn{7}{c}{Total Exposure Time (sec)}\\
            		&           		&     		&(mag)	&           		&             	&\multicolumn{3}{c}{X-shooter} &&\multicolumn{3}{c}{NIRI}\\
\cline{7-9}
\cline{11-13}
\rule{0pt}{2.5ex} 
            		&			&		&    		&			&              		&UVB	&VIS   &NIR		&&$J$ grism&$H$ grism&$K$ grism\\

\hline

0038$-$226	&LHS\,1126	&DQp	&14.3	&NIRI		&2008 Sep 09	&		&		&		&&200	&		&\\
			&			&		&       	&          		&2008 Oct 18 	&		&		&		&&		&		&800\\
			&			&		&        	&          		&2008 Nov 11 	&		&		&		&&		&320		&\\
			&			&		&        	&X-shooter 	&2016 Jul 09 	&1540	&1180	&1620	&&		&		&\\
0341+182		&LHS\,179 	&DQ		&15.1    	&NIRI      		&2008 Sep 13	&		&		&		&&		&800		&1400\\
			&			&		&        	&         		&2008 Sep 24 	&		&		&		&&800	&		&\\
0435$-$088 	&L879-14		&DQ		&13.6    	&NIRI		&2008 Sep 23 	&		&		&		&&200	&180		&240\\
0548$-$001 	&G99-37		&DQ		&14.4    	&X-shooter	&2016 Oct 31	&2200	&1840	&2280	&&		&		&\\
1008+290  	&LHS\,2229	&DQp	&16.5    	&X-shooter	&2017 Feb 01 	&4400	&3680	&4560	&&		&		&\\
1036$-$204 	&LHS\,2293	&DQp	&15.8    	&X-shooter	&2016 Jun 26 	&2740	&2380	&2820	&&		&		&\\
1043$-$188	&LHS\,290	&DQp	&15.4    	&NIRI      		&2008 Nov 11 	&		&		&		&&1040	&		&\\
			&			&		&        	&          		&2008 Nov 16	&		&		&		&&		&1040	&1760\\
			&			&		&        	&X-shooter 	&2017 Mar 23	&2740	&2380	&2820	&&		&		&\\
2008$-$600	&			&DC		&15.6    	&X-shooter 	&2016 Aug 02 	&2740	&2380	&2820	&&		&		&\\
2154$-$512 	&GJ\,841B	&DQ		&14.5    	&X-shooter	&2016 Jul 02 	&1540	&1180	&1620	&&		&		&\\

\hline

\end{tabular}
\end{table*}

The nature of the DQp stars -- specifically their spectral features -- has been debated since the first few discoveries \citep{liebert1978,ruiz1988}.  Although the absorption features range from subtle to enormous \citep[e.g.\ 0038$-$226 vs.\ 1036$-$204;][]{schmidt1995}, it was argued early on that the observed bands are consistent with pressure-shifted Swan bands in a high density atmosphere \citep{liebert1983}.  High density atmospheres were discounted by some models \citep{bergeron1994}, and subsequently magnetism or the C$_2$H molecule were suggested as possibly responsible \citep{schmidt1995}.  Atmospheric hydrogen is plausible for DQp stars, based on a possible dearth of helium-rich atmospheres for $T_{\rm eff}<6000$\,K in studies of chemical evolution of the local white dwarf population \citep{saumon2014,caron2023}.  However, it has been demonstrated that no other molecule but C$_2$ can be responsible for the observed features \citep{hall2008a}.  Similarly, strong magnetism is only observed in a subset of DQp white dwarfs, and thus cannot be responsible for the spectral class as a whole \citep{schmidt1999}.

Pressure distortion is now the accepted model for the blue-shifted centroids and minima in the observed C$_2$ bands of DQp white dwarfs, and such calculations have been successfully incorporated into theoretical spectra. Quantum mechanical calculations demonstrated that when C$_2$ is compressed in dense helium, the electronic transition energy of the Swan bands increases, leading to the observed blueward shift \citep{kowalski2010}.  The appearance of the distortion in cooler stars is a consequence of their helium-dominated atmospheres approaching, and then exceeding, the density of water. The latter results from the decreasing number of free electrons from helium and carbon ionization at lower temperatures \citep{dufour2005,koester2006b}.  This pressure distortion model has been used to reproduce the Swan bands in several DQp stars \citep{kowalski2010,blouin2019b}, although the entire spectral energy distributions of these cool white dwarfs could not be reproduced even with the advanced atmosphere models, particularly in the near-infrared \citep{blouin2019b,blouin2024}.

\begin{figure*}
\includegraphics[width=\linewidth]{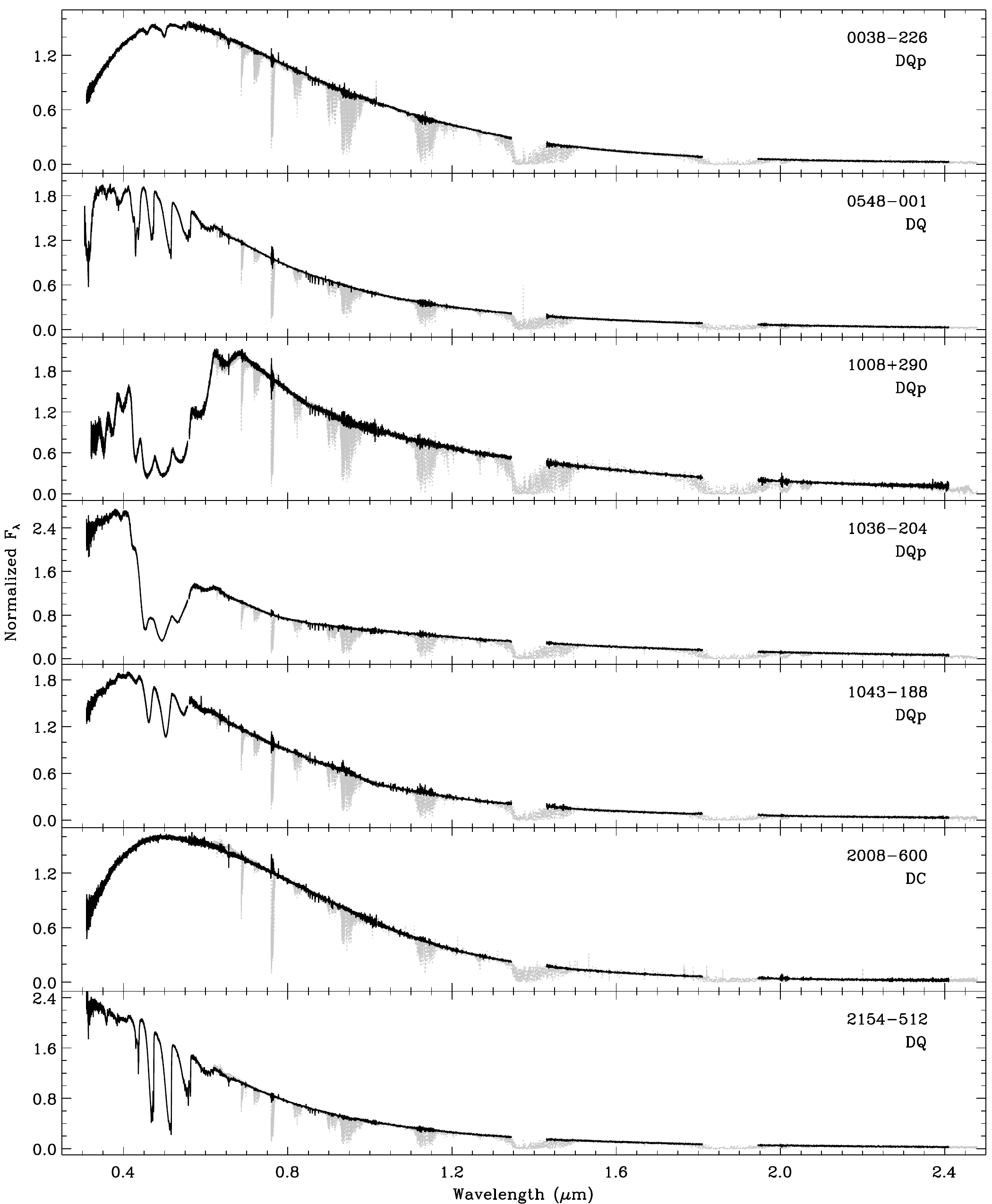}
\vskip 0 pt
\caption{All reduced, combined, and telluric-corrected X-shooter spectra, plotted in black in order of right ascension, and displayed at full spectral resolution without smoothing or binning.  The uncorrected spectra are shown in grey and exhibit atmospheric absorption features in many regions.  While the corrected spectra contain significant residuals in regions of telluric absorption, and where the standard stars contain Paschen and Balmer lines, the overall shape of the spectral energy distributions are recovered at high fidelity.  The faintest target is 1008+290 and its spectrum exhibits the most significant effects of telluric absorption and residuals.}
\label{allxs}
\end{figure*}

This study attempts a comprehensive characterization of cool DQ and DQp stellar atmospheres by providing consistent model fits to multi-wavelength spectra spanning the blue optical through the near-infrared.  The structure of the paper is as follows.  Section~2 details the target selection, observations, and data reduction, and Section~3 describes the fitting of these spectra using state-of-the-art atmosphere models. Section~4 contains the results and some in-depth discussion of individual stars and modeling results, and Section~5 contains the conclusions.

\section{Observations and data}

Observations for this study were carried out at two facilities, the European Southern Observatory Very Large Telescope, and the Gemini North Observatory.  Eight white dwarfs of spectral type DQ or DQp, four of each, were selected for multi-wavelength characterization based primarily on their near-infrared brightness to achieve relatively high signal-to-noise (S/N) spectroscopy at these wavelengths, but also to have cool effective temperatures spanning the transition from DQ to DQp spectral morphology.  A ninth, ancillary target of spectral type DC was also observed as a comparison target lacking carbon features but having broadly similar stellar parameters, and photometry indicative of infrared flux suppression.  The targets and observing details are summarized in Table~\ref{obs}.

\begin{figure*}
\includegraphics[width=\linewidth]{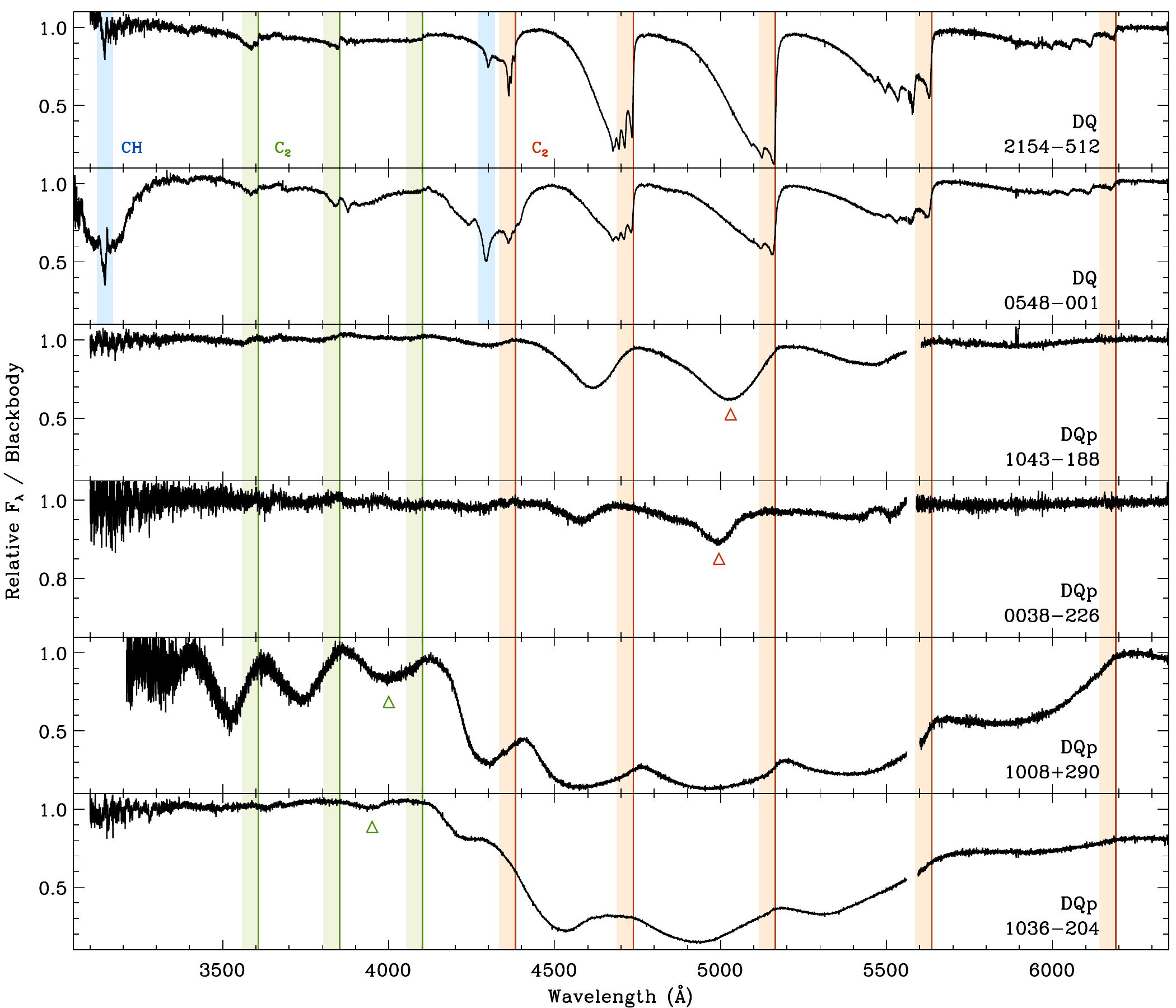}
\vskip 0 pt
\caption{Close-up X-shooter spectra, in order of decreasing $T_{\rm eff}$ from top to bottom.  Each spectrum is divided by a blackbody with a temperature such that the resulting continuum is relatively flat, and normalized to 1.0.  For these two DQ and four DQp stars, the Deslanders-d'Azambuja (green) and Swan bands (orange) of C$_2$ are shaded, with darker lines of the appropriate hue at the nominal, unshifted bandhead wavelengths.  In the DQp white dwarfs, easily visible examples of the shifted bands are marked for both sets of C$_2$ transitions with colored triangles. In the two DQ stars, CH (blue) is detected near 3145\,\AA~and in the G band around 4300\,\AA~\citep{hollands2022}.}
\label{allc2}
\end{figure*}

\subsection{X-shooter echelle spectroscopy}

Seven targets were observed with the triple-arm, echelle spectrograph X-shooter \citep{vernet2011}.  The instrument was chosen based on its extensive wavelength coverage from 3000\,\AA\ to 2.5\,$\upmu$m, where observational goals included the detection of carbon molecular features not previously observed in white dwarfs; namely the Deslanders-D’Azambuja bands in the blue-ultraviolet and and the Ballik-Ramsay bands in the near-infrared.

The observations were carried out in service mode in programs 097.D-0064 and 098.D-0059, with the telescope nodding 5\,arcsec along the slit to enable accurate sky subtraction in the near-infrared arm of the spectrograph.  For this purpose, one or more ABBA exposure sequences were carried out for each star in all three arms.  The UVB and VIS detectors were employed in high-gain, slow-readout mode, with no binning in the spatial direction, and two-pixel binning in the dispersion direction.  Spectrographic slits of sizes 1.0, 0.9, and 0.9\,arcsec were utilized for the UBV, VIS, and NIR arms, respectively, for nominal resolving powers of 5400, 8900, and 5600.  

The X-shooter data were processed using version 2.9.3 of the reduction pipeline, following standard procedures, using the {\sc esorex} reduction software version 3.12.  Each spectrum was extracted and wavelength-calibrated using default settings, with the exception that the position of the target spectrum was localized on the slit using Gaussian fitting and allowing for some tilt.  The pipeline produces a single spectrum for each star in each arm, which is an average of the individual exposures.  The data were sensitivity-calibrated using spectrophotometric standard star observations taken on the same night.  The resulting spectra typically have S/N $>100$ in the UVB and VIS arms, and S/N $>20$ in the NIR arm.

Further processing was carried out to achieve the most accurate spectral morphology across all wavelengths.  Telluric correction was enabled by the observation of early-type standard stars, provided as calibrations by the observatory.  The minimization of telluric residuals was performed using {\sc xtellcor\_general} from within the {\sc spextool} package \citep{cushing2004}, and was done for both the red optical portion of the spectra as well as the infrared region.  Overall, while many telluric features were removed or minimized, there remain some residuals that were challenging to eliminate.  The resulting spectral slopes in each arm are accurate, but not necessarily the fluxes.  The dichroic crossover regions between the spectrograph arms (near 5560 and 10\,200\,\AA) were examined, and the VIS and NIR spectra offset so that the fluxes across these narrow wavelength ranges are consistent, after which the entire spectrum was normalized to a mean of one.  All the resulting X-shooter spectra are shown in Figure~\ref{allxs}, with a closer view of the detected carbon features in Figure~\ref{allc2}.

\subsection{NIRI grism spectroscopy}

Near-infrared spectroscopic observations for four targets (two in common with X-shooter) were obtained at Gemini Observatory North, using the Near-infrared Imager and Spectrometer \citep[NIRI;][]{hodapp2003}, under program GN-2008B-Q-110. NIRI was used with the f/6 camera, providing a plate scale of 0.12\,arcsec\,pixel$^{-1}$, and a 6-pixel wide or 0.7\,arcsec slit. For each star, three grism and slit combinations were used: the $J$-grism with the blue slit provided wavelength coverage 0.99--1.35\,$\upmu$m at resolving power $R\approx480$; the $H$- and $K$-grisms with the centered slit covered 1.43--1.96\,$\upmu$m and 1.90--2.49\,$\upmu$m, respectively, both with $R\approx520$.  Targets were nodded 6\,arcsec along the slit during observation, in the standard ABBA pattern, repeated once or twice depending on target brightness.

For each star with each grism, a late-A to mid-F type star with $J\approx8$\,mag was used as a standard to remove the effects of the terrestrial atmosphere, with H\,{\sc i} recombination lines in their spectra removed artificially prior to taking the ratio of target to standard star.  Lamps in the on-instrument calibration unit were used to provide flat fielding and wavelength calibration. The data were reduced in the standard way using a combination of {\sc iraf} and {\sc starlink figaro} package routines.

\subsection{Ancillary spectral data}

Optical spectra for the NIRI targets 0341+182 and 0435$-$088 \citep{bergeron1997} were accessed from the Montreal White Dwarf Database\footnote{\url{https://montrealwhitedwarfdatabase.org}}.  As there are no overlapping regions between optical and infrared wavelengths for these two targets, some manual adjustments are necessary using broad-band photometry and guided by the modeling.

\section{Atmospheric model fitting}

\begin{figure*}
\includegraphics[width=\linewidth]{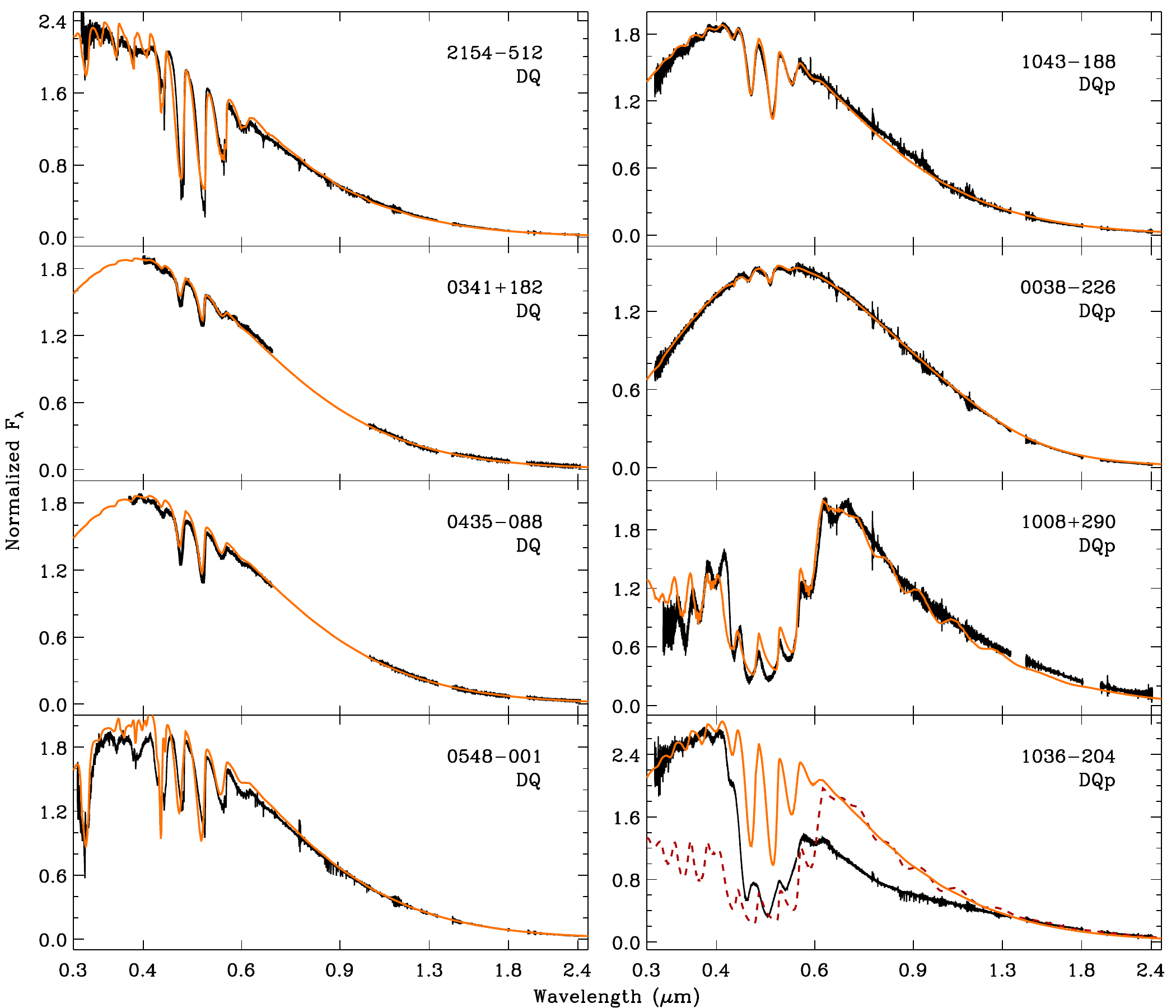}
\caption{Observed and best synthetic spectra of DQ and DQp stars, in order of decreasing $T_{\rm eff}$ from top left to bottom right.  The data are plotted in black and overplotted with models in solid orange, with the exception of 1036$-$204; for this star the solid and dashed lines indicates the warmer and cooler model fits from Table \ref{results}.}\label{allmods}
\end{figure*}

The spectral analysis for the sample of observed stars is performed by deriving a fine grid of atmosphere models using $T_{\rm eff}$, [H/He], and [C/He] as parameters. The models are computed by a custom white dwarf atmosphere code that includes all previous improvements in physics, chemistry, radiative transfer, and absorption in dense helium \citep[e.g.][]{kowalski2004,kowalski2007,blouin2017}.  

The models include the distortion of C$_2$ in dense helium \citep{kowalski2010}, the Ly\,$\upalpha$ red wing absorption by H$\rm _2$-He collisions \citep{kowalski2006b}, and the near-infrared absorption from helium atoms \citep{kowalski2014}, all essential for accurate model fitting.  In previous studies \citep[e.g.][]{kowalski2010}, C$_2$ opacities from \citet{freedman2008} have been used but do not perfectly reproduce the C$_2$ bands strengths \citep[in particular that of $\Delta\nu=+2$ band;][]{dufour2005}.  Here, new C$_2$ opacities are computed using published oscillator strengths and Franck-Condon factors \citep{clementi1960,nicholls1965,brooke2013}, and include the following absorption bands: Mulliken (2300--2400\,\AA), Deslanders-d'Azambuja (3400--4100\,\AA), Swan (4300--7900\,\AA), Phillips (0.77--1.21\,$\upmu$m) and Ballik-Ramsay (1.20--2.50\,$\upmu$m). 

A key aspect of the modeling is to correctly describe the ionization equilibrium in dense helium, because free electrons contribute directly and indirectly to the opacity.  The ionization fraction depends on the atmospheric density, which determines the strength of the near-infrared absorption, including collision-induced absorption (CIA) from hydrogen or helium fluids \citep{kowalski2014,blouin2017}, which are constrained here by analyzing the observed infrared spectra.

Bands of CH are detected in two of the objects studied here (see Figure~\ref{allc2}), and for this reason the molecule is included in the modeling by applying the absorption cross-sections for the A–X, B–X, and C–X transitions and the CH partition function of \citet{jorgensen1996}.  It has been verified that these result in a G-band strength that is consistent with those derived using independent models of cool DQ stars \citep{blouin2019a}.  To fit the optical C$_2$ bands and derive [C/He] abundances, synthetic photometry is also computed for artificial bandpasses in the ranges 5300--5500\,\AA\ and 5050--5150\,\AA, each covering one of the C$_2$ Swan absorption bands.  Photometry in these artificial bands is integrated from optical spectra that are rescaled to match the observed flux in the $V$ band.

Using the model spectra, synthetic photometry is computed in common broadband filters, including $UBVRIJHK$, $ugriz$, and {\em Gaia} photometry.  The atmospheric parameters are derived by a least squares fit between the synthetic and observed photometric spectral energy distributions obtained from existing photometric surveys accessed from the Montreal White Dwarf Database (see Section 2.3 footnote), further refining the parameters by fitting the synthetic to the observed spectra.

Initially $\log\,g=8.0$ is assumed for all stars. This reduces the parameter space to $T_{\rm eff}$, [H/He], and [C/He], which primarily determine the shape of spectrum. Once the atmospheric parameters for each star are determined, synthetic spectra are computed and the [C/He] ratio is further (slightly) adjusted to best match the observed C$_2$ band strengths. The value of $\log\,g$ is also refined in subsequent iterations of the least squares fit, and checked for consistency with the radius and mass values determined using evolutionary models and the known distance to each star.  Because all the white dwarfs studied here have precise {\em Gaia} parallax measurements, the combination of synthetic and observed {\em Gaia} photometry directly constrains the radius of each star.  Using the atmospheric parameters together with evolutionary models for pure helium atmospheres \citep{fontaine2001}, the masses of the DQ and DQp stars are derived.

\section{Results and discussion}

This section contextualizes the results, highlights the significant spectral features detected, and briefly summarizes the modeling outcomes for the DQ stars.  Individual DQp stars are discussed in more detail, in particular 0038$-$226 and the two coolest stars in the sample.  Lastly there is a discussion of the implications based on the modeling and the overall properties of these stars.  The modeled atmospheric and stellar parameters are given in Table~\ref{results} for the nine stars studied here - four DQ, four DQp, and the carbon-free DC star which acts as a comparison object.  Figure~\ref{allmods} compares the best-fit synthetic spectra to the observational data for the eight DQ(p) stars.  

Table~\ref{results} also includes prior modeling efforts for the stars studied here, and the new results are consistent with the earlier work.  The table provides notional uncertainties for the $T_{\rm eff}$ and $\log\,g$ values determined here, except for 1036$-$204, which remains challenging (see Section~4).  For all other stars the uncertainty in $\log\,g$ is 0.03\,dex, and primarily due to the uncertainty in $T_{\rm eff}$, with a small contribution from the uncertainty in the {\em Gaia} parallax and photometry.  The uncertainty in $T_{\rm eff}$ is estimated by varying the value slightly and performing least squares fits of the resulting models to the measured fluxes, adopting $1\upsigma$ confidence.  It is important to note that the errors given in Table~\ref{results} are merely formal errors and do not include systematic uncertainties, such as deficiencies in the models, or the impact of echelle multi-order data compared to single-order spectroscopy.  It is clear there is not yet a complete understanding of cool atmospheres at the extreme densities considered here. The true uncertainty in $T_{\rm eff}$ is estimated to be around 100\,K.

\subsection{Spectral features}
 
Distinct spectral features are only detected in the near-ultraviolet and optical wavelength range for all targets.  These are highlighted in a closer view of the relevant X-shooter data in Figure~\ref{allc2}, covering the entirety of both the Deslanders-d'Azambuja and Swan C$_2$ molecular bands, as well as the 3145\,\AA\ and G band of CH.  As shown in Figure~\ref{allxs}, the observed near-infrared spectra of all program stars are featureless, and there is no indication of the C$_2$ Ballik-Ramsay bands in any DQ or DQp white dwarf. 

\begin{figure}
\includegraphics[width=\columnwidth]{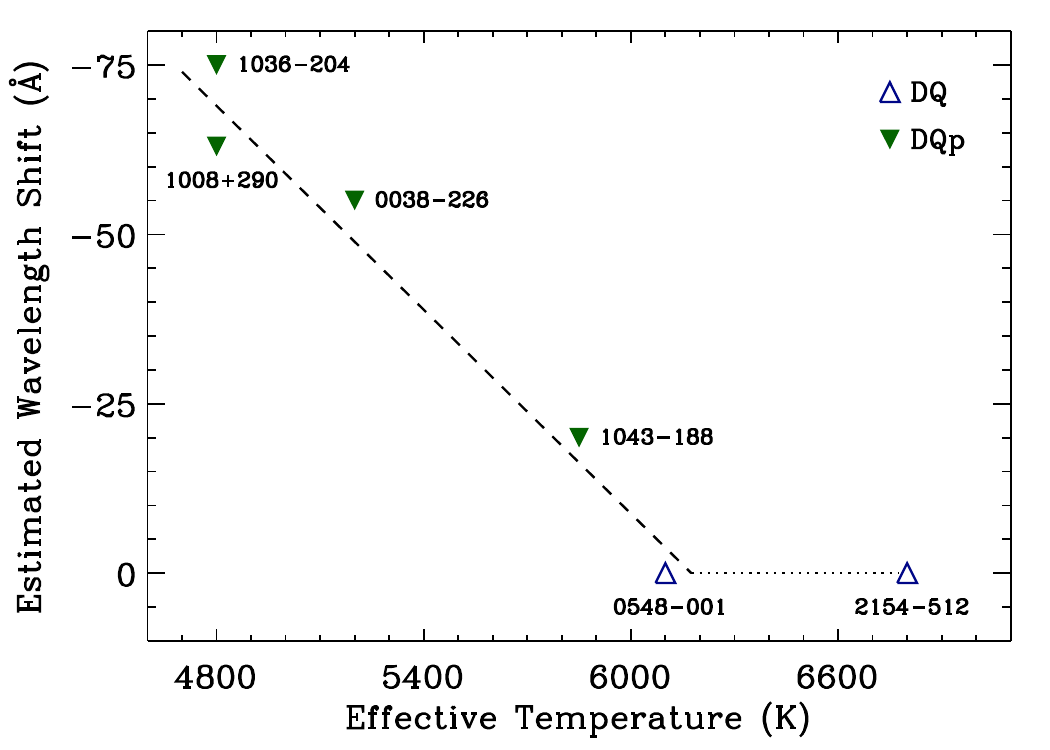}
\caption{The estimated wavelength shift in the Deslanders-d'Azambuja bands in the spectra of DQ and DQp stars as a function of effective temperature.  The observed shifts were calculated as the difference between the nominal, unshifted bandhead wavelength and the observed minimum of the band, using the average of the 3607 and 3852\,\AA\ bands for each star.  The shifts follow a linear dependence with effective temperature for $T_{\rm eff}<6200$\,K, shown here by a dashed line.  Based on Figure~\ref{allmods}, the shifts are reasonably reproduced by the Swan band distortion model \citep{kowalski2010}.}
\label{blueshift}
\end{figure}

The well-known distortion of the Swan bands is visible as blueward shifts of the absorption feature minima in the DQp white dwarfs.  Notably, the X-shooter data also contain analogous shifts for the Deslanders-d'Azambuja bands -- observed here for the first time (unshifted bands have been previously detected in DQ stars; \citealt{dufour2005,blouin2019b}).  The models in Figure~\ref{allmods} reproduce the observed distortion of the Deslanders-d'Azambuja bands reasonably well, where the predictions are based on published distortion models for the Swan bands \citep{kowalski2010}.  However, this could be coincidence, as the size of density-induced distortion has not been established for these bands, which arise from transitions $\rm A ^1\Pi_u$,$\rm C ^1\Pi_g$, whereas those involved in the Swan bands are $\rm d ^3\Pi_u$,$\rm a ^3\Pi_g$ \citep{nicholls1965}.  As shown in Figure~\ref{blueshift}, unlike the Swan bands, the shift of the minima of the Deslanders-d'Azambuja bands is not constant but increases with decreasing effective temperature. This trend appears to follow atmospheric density and provides another constraint on effective temperature for the coolest stars in the sample.

\begin{table}
\begin{center}
\caption{Atmospheric model parameters of DQ and DQp stars derived here and in prior work, ordered by decreasing effective temperature.  Abundances are [X/He] $= \log{\rm (X/He)}$.  The quoted uncertainties are formal errors and do not include systematics, and are thus underestimated (see Section~4).}\label{results}
\setlength{\tabcolsep}{4pt}
\begin{tabular}{lllrrcc}

\hline

WD      		&$T_{\rm eff}$  		&$\log\,g$  	&[C/He]     &[H/He]  	&Mass          	&Ref.\\
        			&(K)            		&(cgs)      		&           	&        	&(M$_\odot$)   	&\\
\hline
\multicolumn{6}{l}{DQ stars:}\\

2154$-$512  	&$6750\pm75$		&$8.05\pm0.03 $     	&$-$5.0	&$-$4.8  	&0.60          &1\\
            		&$6549\pm41$       	&$7.97\pm0.02$       	&           	&	     	&0.55          &2\\    
0341+182    	&$6466\pm85 $       	&$8.00\pm0.03 $      	&$-$6.6   	&$<-$5.3 	&0.58          &1\\
            		&$6837\pm48$       	&$8.10\pm0.02$       	&           	&  	     	&0.63          &2\\
            		&$6515\pm60$       	&$7.97\pm0.03$       	&$-$6.5    	& 	     	&0.56          &3\\
            		&$6548\pm29$       	&$7.98\pm0.01$       	&$-$6.4    	& 	     	&0.56          &4\\
0435$-$088  	&$6248\pm60 $       	&$7.94\pm0.03$       	&$-$6.6    	&$<-$6.1 	&0.54          &1\\
            		&$6601\pm38$       	&$8.05\pm0.02$       	&           	&	     	&0.60          &2\\
            		&$6395\pm30$       	&$7.99\pm0.01$       	&$-$6.5    	& 	     	&0.55          &3\\
            		&$6275\pm7$       	&$7.91\pm0.02$     	&$-$6.4    	&	     	&0.52          &4\\
0548$-$001  	&$6057\pm49$     	&$8.18\pm0.03$       	&$-$6.5    	&$-$5.0	&0.69          &1\\
            		&$6053\pm30$       	&$8.12\pm0.02$       	&           	&        	&0.65          &2\\
            		&$6080\pm45$      	&$8.15\pm0.02$	&$-$6.6    	&$-$4.3	&0.66          &3\\ 
            		&$6235\pm4$       	&$8.20\pm0.01$     	&$-$6.5    	&	     	&0.70          &4\\
            		&$6228\pm4$       	&$8.16\pm0.01$     	&$-$6.1    	&	     	&0.67          &5\\
\hline

\multicolumn{6}{l}{DQp stars:}\\

1043$-$188  	&$5886\pm52$       	&$7.99\pm0.03$    	&$-$7.0    	&$<-$6.9 	&0.57         &1\\
            		&$5736\pm50$       	&$7.88\pm0.03$       	&           	&	     	&0.50         &2\\
            		&$5832\pm86$       	&$7.86\pm0.05$       	&$-$6.7   	& 	     	&0.49         &3\\ 
            		&$5387\pm17$       	&$7.59\pm0.09$     	&$-$7.2    	& 	     	&0.35         &4\\
0038$-$226 	&$5235\pm55$       	&$7.97\pm0.03$       	&$-$8.9    	&$-$6.8  	&0.55         &1\\
            		&$5368\pm21$       	&$7.99\pm0.01$       	&           	&	     	&0.56         &2\\
            		&$5210\pm60$       	&$7.91\pm0.04$       	&$-$8.4    	& 	     	&0.51         &3\\
            		&$5307\pm7$       	&$7.95\pm0.03$     	&$-$8.0    	& 	     	&0.54         &4\\     
            		&$5262\pm23$       	&$7.99\pm0.00$       	&          	&	     	&0.56         &5\\
1008+290    	&$4778\pm84$       	&$8.37\pm0.03$       	&$-$6.6    	&$<-$7.0 	&0.81         &1\\
            		&$4335\pm165$       &$8.21\pm0.09$       	&$-$6.8    	&	     	&0.70         &3\\
            		&$4595\pm18$       	&$8.27\pm0.01$       	&$-$6.7   	&	     	&0.74         &4\\ 
1036$-$204  	&4800         		&8.08         		&$-$6.6    	&$<-$7.5 	&0.62         &1\\
            		&5800         		&8.59  	    		&$-$6.6    	&	     	&0.96         &1\\
            		&$5754\pm54$       	&$8.56\pm0.02$	&           	&        	&0.94         &2\\
            		&$4530\pm215$     	&$8.07\pm0.12$	&$-$7.2    	&        	&0.61         &3\\  
            		&$5362\pm9$       	&$8.45\pm0.03$	&$-$6.6    	&	     	&0.87         &4\\ 
\hline
\multicolumn{6}{l}{DC star:}\\

2008$-$600    	&$5336\pm43$       	&$8.10\pm0.03$&\multicolumn{2}{c}{pure He}	&0.64&1\\

\hline

\end{tabular}
\end{center}
{\em References:} (1) This work; (2) \citet{obrien2024}; (3) \citet{blouin2019a,blouin2019b}; (4) \citet{caron2023}; (5) \citet{kilic2025}.
\end{table}

\subsection{Results for stars with \texorpdfstring{$T_{\rm eff}>5500$\,K}{Teff > 5500K}}

Figure~\ref{allc2} shows that the CH bands are clearly detected in the two DQ stars with X-shooter data: 0548$-$001 and 2154$-$512.  These detections, as well as upper limits, constrain both the carbon and hydrogen abundances, where the latter is also constrained by the strength of the C$_2$ bands. This is because the addition of hydrogen raises the atmospheric opacity, decreases density and pressure, and thereby increases the amount of carbon required to reproduce the carbon bands.  For example, if it were the case that [H/He] $=-3$ in 0341+182, the carbon abundance would increase by a factor of three.  All four DQ stars are found to have [H/He] $ < -4$.

There have been several studies showing that atmospheric models can accurately reproduce optical through near-infrared photometry, and spectra in the Swan bands region of DQ stars \citep[e.g.][]{dufour2005,giammichele2012}.  Previous modeling work consistently finds [C/He] ratios that decrease with decreasing effective temperature \citep{kowalski2013,blouin2019b, koester2019}, which is the expected trend based on evolutionary models \citep[e.g.][]{bedard2024a}).  Therefore, similarly good fits are expected and achieved here, in good agreement with previous studies, with some minor differences as seen in Table~\ref{results}.

The warmest DQp star in the sample is 1043$-$188, and the model fit shown in Figure~\ref{allmods} for this source is excellent.  The determined effective temperature is in agreement with two independent analyses, where the mass found here is 10\,per cent higher \citep{blouin2019b,obrien2024}.  However the results here differ from those of \citet{caron2023}, where a mass of 0.35\,M$_\odot$ is found, which is too low for single star evolution in a Hubble time.

\subsection{\texorpdfstring{0038$-$226}{0038-226}}

This white dwarf has additional data that includes near-ultraviolet spectroscopy with the Faint Object Spectrograph on the {\em Hubble Space Telescope} \citep{wolff2002}, and mid-infrared photometry from 3 to 8\,$\upmu$m using the Infrared Array Camera on the {\em Spitzer Space Telescope} \citep{kilic2008}.  There are also {\em James Webb Space Telescope} spectra across the near- and mid-infrared \citep{blouin2024}, which exhibit no features and are consistent with the {\em Spitzer} photometry.  The ultraviolet data permit a more accurate derivation of the hydrogen content of 0038$-$226, and the mid-infrared data provide an additional test of the CIA opacity.

For this object, the modeling relies on the dense helium effects on the Ly\,$\upalpha$ red wing opacity \citep{kowalski2006b}, the C$_2$ Swan band distortion \citep{kowalski2010}, and the He CIA opacity \citep{kowalski2014}.  Most prior modeling of this star assigned the infrared flux suppression to a strong H$_2$-He CIA \citep[e.g.][]{giammichele2012,blouin2019a}, but these models are inconsistent with the featureless infrared data \citep{kilic2008,blouin2024}. The upper limit on CH in 0038$-$226 yields [H/He] $<-6.7$, and is independently corroborated by a fit to the Ly\,$\upalpha$ red wing profile that yields [H/He] $=-6.8$.  This indicates that He CIA is responsible for the shape of the infrared flux suppression.

\begin{figure}
\includegraphics[width=\columnwidth]{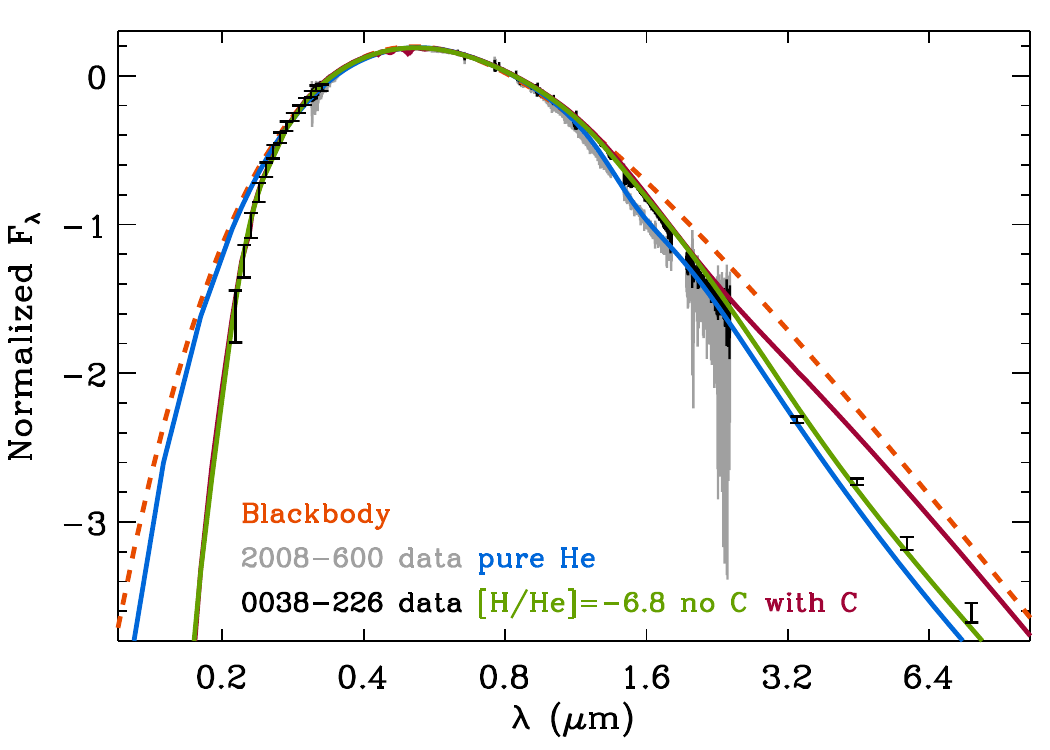}
\caption{A comparison of atmospheric models and data for 0038$-$226 (DQp) and 2008$-$600 (DC), which have similar $T_{\rm eff}$ and $\log\,g$.  The data for 0038$-$226 are plotted in black, including ultraviolet and infrared data from {\em Hubble} and {\em Spitzer}, respectively, and are fitted with models that have trace hydrogen but no carbon (green), and trace carbon (magenta). The data for 2008$-$600 are plotted in grey, and overplotted with a pure He model (blue).  A dashed orange line shows a blackbody for comparison, emphasizing the differing degrees of infrared flux suppression in both stars.  A key result is that the infrared flux suppression of both stars can be reproduced by hydrogen-poor models using only He CIA, but, with the introduction of trace carbon for 0038$-$226, the model deviates from the observations in the mid-infrared.}
\label{comp2008}
\end{figure}

The data for this star constrains the highly uncertain helium ionization equilibrium in white dwarf atmospheres \citep{kowalski2007,blouin2024}.  The helium ionization energy provided by the chemical models are rescaled by a factor of 0.375, so that the near-infrared spectrum is well reproduced. The resulting effective ionization energy of helium for the photosphere of 0038$-$226 is 20.5\,eV at a density of 1.2\,g\,cm$^{-3}$ and $T_{\rm eff}=5900$\,K.  This is consistent with the ionization energy in fluid helium computed with advanced quantum mechanical calculations \citep[][figure~10]{kowalski2007}. This may indicate that direct ionization and not entropic effects (which govern the ionization equilibrium in a chemical model) determines the number of free electrons in fluid helium.

Support for this treatment of the helium ionization, and for the importance of He CIA, comes from the fit to the DC white dwarf 2008$-$600. This star has a similar $T_{\rm eff}$ and surface gravity to 0038$-$226, but has a pure He atmosphere.  Figure~\ref{comp2008} shows the fits to 0038$-$226 and 2008$-$600 simultaneously using the same axes.  Notably, the CIA opacity is stronger in the infrared for the pure He DC white dwarf because the trace carbon and hydrogen in 0038$-$226 make the atmosphere more opaque and less dense.

Figure~\ref{allmods} shows that the model with trace carbon does an excellent job of reproducing the near-ultraviolet to near-infrared spectrum of 0038$-$226, including the distorted C$_2$ Swan bands.  However Figure~\ref{comp2008} shows that the mid-infrared {\em Spitzer} data are better reproduced by a carbon-free atmosphere.  Further exploration of this is beyond the scope of the current paper, however preliminary work suggests that the temperature profile of the atmosphere may be changed by the addition of carbon, which could lead to changes in the mid-infrared.  It is also possible that carbon is not uniformly distributed throughout the atmosphere, which would also change the temperature profile and the spectral energy distribution.

\begin{figure}
\includegraphics[width=\columnwidth]{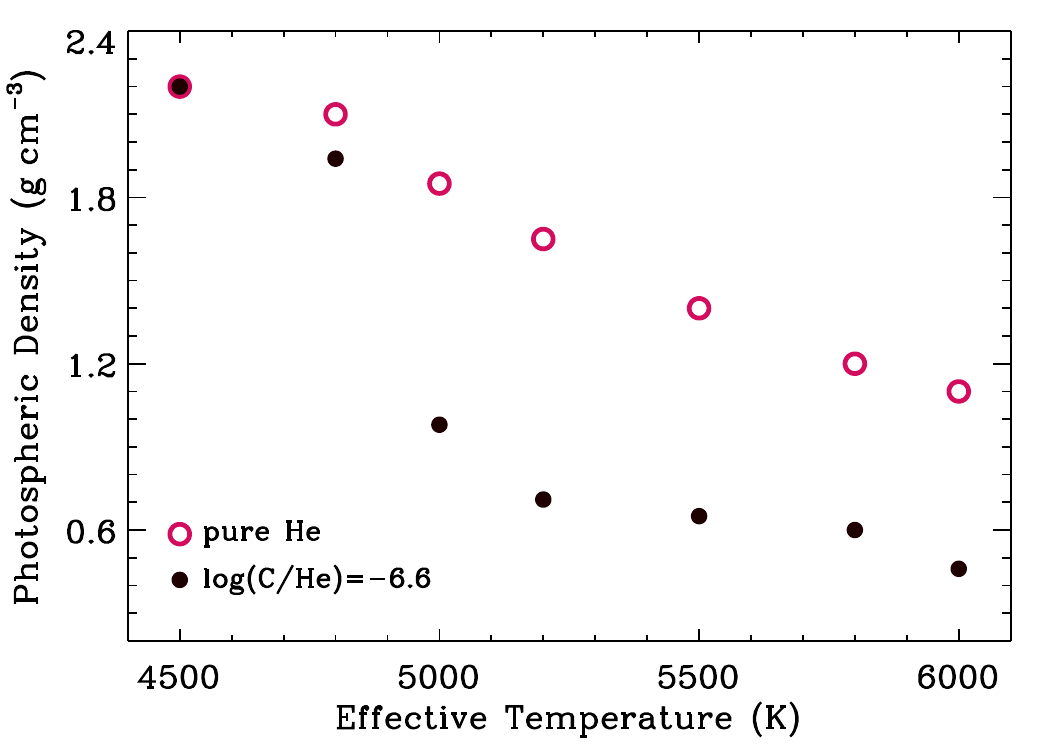}
\caption{The photospheric density of helium atmosphere white dwarfs with [C/He] = $-6.6$ obtained in the applied atmospheric models \citep[cf.][figure~4]{blouin2019b}.}\label{density}
\end{figure}

\subsection{1008+290 and \texorpdfstring{1036$-$204}{1036-204}}

The white dwarfs 1008+290 and 1036$-$204 are the two coolest objects in the sample.  Their spectral morphologies are drastic departures from a simple continuum, and their data are not as well reproduced by the models as the other DQ and DQp stars in Figure~\ref{allmods}.  Although the C$_2$ band wavelength shifts are broadly reproduced for both stars, as in previous studies, these spectra continue to present modeling challenges \citep[e.g.][]{blouin2019a}.

In 1008+290, the modeling qualitatively reproduces the X-shooter data, including the variegated shapes and depths of nearly all the C$_2$ bands, and the jagged transition to the continuum at longer wavelengths.  The spectral regions for which the model struggles are at the bluest X-shooter wavelengths, where the model over-predicts the quickly decreasing flux towards the ultraviolet (i.e.\ $NUV-u=3.9$\,AB\,mag), and the observed flat continuum in the near-infrared.  This is the only target in which the relatively weak, near-infrared Ballik-Ramsay C$_2$ bands are predicted to be observed (but are not).  Infrared flux suppression in this star may play a role in reducing the depth of these C$_2$ features so they are not apparent in the data, where, unfortunately, this is the faintest star in the sample and has the lowest S/N.

In 1036$-$204, it has been previously noted that the spectral energy distribution outside of the deep and blended Swan bands does not have a consistent slope; the ultraviolet-blue region can be approximated by a 7500\,K blackbody, while the red to near-infrared follows a 4300\,K blackbody \citep{farihi2009b}.  This dualistic behavior can be seen in Figure~\ref{allc2}, where continuum normalization with a single blackbody leads to uneven levels in the blue and red, and in Figure~\ref{allmods} where the changing slope cannot be fully reproduced with the atmosphere models used here.  The detected Deslanders-d'Azambuja band distortion is stronger than that observed in 1008+290, and may indicate that this extremely cool target has a helium-rich atmosphere that is even more dense than calculated here.  

 Figure~\ref{allmods} shows that the 5800\,K solution for 1036$-$204 suggests there is a missing opacity source in the 0.4--1.3\,$\upmu$m region.  As illustrated in Figure~\ref{density}, the preferred, 4800\,K solution for this star indicates an extreme density approaching 2\,g\,cm$^{-3}$, where opacities have not been calculated.  For example, He CIA opacity has only been modeled for densities up to 0.8\,g\,cm$^{-3}$, and studies have shown that increasing density can distort and blueshift CIA absorption \citep{kowalski2014,blouin2017}.  In addition, there are short-wavelength peaks in the H$_2$-He CIA opacity \citep[e.g.\ fig.~2][]{borysow1997} which may be important for this object.  These issues underscore the need for modeling at higher densities.

\subsection{Stellar mass and magnetism}

 Analyses of larger samples of DQ stars derive masses that are smaller than the typical white dwarf mass by $\sim$0.05\,M$_\odot$ \citep{coutu2019,koester2019,blouin2019b}, furthermore the mass distributions of DQ stars appear distinct \citep{bedard2022}. While it has been suggested that atmospheric modeling improvements may reconcile the apparent mass deficit, it is also possible that the evolutionary paths for DQ stars and their cooler DQp descendants are different \citep{farihi2024,bedard2024b}.  The sample studied here is too small to constrain such a trend; overall and particularly for the warmer stars, the masses appear broadly consistent with the typical white dwarf mass of 0.60\,M$_\odot$ \citep[e.g.][]{genest2019}.

 The two coolest stars in the sample may depart from this behavior.  A relatively high mass is determined for 1008+290, and a high mass cannot be excluded for 1036$-$204.  If indicative, this would suggest that these two stars are merger remnants, or descendants of B-type main-sequence stars.  However, given the difficulties in modeling the coolest DQp stars, these high masses should be treated with caution.

 Four of the stars in the sample are detectably magnetic via spectropolarimetry.  The DQ white dwarfs 0548$-$001 and 2154$-$512 have fields between 1 and 5\,MG \citep{angel1971,vornanen2010}, and the two coolest DQp stars 1008+290 and 1036$-$204 both have strong fields on the order of 100\,MG \citep{liebert1978,schmidt1995,schmidt1999}.  The latter may be part of the observed correlation between magnetic field incidence and increasing cooling age for white dwarfs that evolve as single stars \citep{bagnulo2022}, however it may also be the case that DQ stars are more frequently magnetic or exhibit stronger fields than average \citep{bagnulo2021}.  The effects of strong magnetic fields on the C$_2$ molecule and on CIA have not been computed.  While there must be a significant effect on both the absorption features and the stellar continuum -- since polarization is detected both inside and outside the Swan bands -- there is currently no available quantification of these effects for the modeling.  It is noteworthy that the two magnetic DQ stars have excellent model fits to their data.

\subsection{Implications for cool white dwarf evolution}

 The novel modeling presented here finds that the photospheric densities of the coolest stars are $\uprho\approx2$\,g\,cm$^{-3}$, approaching the density of magma \citep{hack2010} and should continue to increase toward cooler effective temperatures as indicated in Figure~\ref{density}.  This fact underscores the need to understand dense fluid effects in modeling efforts.  It is important to be cautious using atmospheric models at $T_{\rm eff}<5000$\,K, where dense, helium fluid effects become important but often have not been derived or tested for validity at $\uprho>1$\,g\,cm$^{-3}$ (e.g.\ the profiles for pure He CIA). These effects are accounted for here in the modeling, with the aforementioned uncertainty in the actual ionization equilibrium and density.

 The results support the interpretation of DQp stars as descendants of the DQ spectral class, with similar masses but with distorted carbon bands \citep{kowalski2010}.  The carbon abundances follow the trend found in previous studies, with [C/He] decreasing by about an order of magnitude per 1000\,K, with no disruption across the DQ to DQp transition \citep{koester2006b,blouin2019b,koester2020}.  Only trace amounts of hydrogen are inferred for the atmospheres of the coolest stars, and for the most part reproduces the broad energy distributions from the ground-based ultraviolet through the near-infrared.  A recent study by \citet{caron2023} finds that hydrogen is needed to describe near-infrared flux suppression in some stars, whereas here it has been demonstrated that He CIA can produce a similar effect. It is thus plausible that infrared-faint white dwarfs have a significantly lower hydrogen content than previously suggested.

 It is conceivable that the stars studied in this small sample are outliers and thus not highly representative of the overall population of cool, carbon-enriched white dwarfs.  The current understanding of the oldest and coolest white dwarfs will likely improve with ongoing, wide-field spectroscopic surveys and future modeling.  Importantly, the best constraints ideally require ultraviolet and near-infrared data, and these wavelength regions are neither currently available nor will be covered in ongoing spectroscopic or photometric surveys that will observe large populations of cool white dwarfs.

\section{Conclusions}

This work has fully characterized four DQ and four DQp stars based on X-shooter, NIRI, and ancillary optical spectra spanning the blue optical to the near-infrared.  The Deslanders-d'Azambuja C$_2$ bands are detected in all the stars with short-wavelength data, and for the DQp stars these display analogous distortions to the previously known blue-shifted Swan bands, with increasing blueward shifts with decreasing effective temperature.  The spectra cover wavelengths for the Ballik-Ramsay C$_2$ band in the near-infrared, where they are predicted to be significant in only one target, but are not detected.  

With one exception, the models have successfully reproduced the overall shapes of the spectral energy distributions, all the observed C$_2$ features in both depth and shape, including the blue shifted, pressure-induced distortions, and infrared flux suppression observed in all the DQp white dwarfs.  All models employed here are hydrogen-poor with [H/He] $<-4$ for the (warmer) DQ stars and [H/He] $<-6$ for the (cooler) DQp stars.  Critically, and based on these trace hydrogen constraints, infrared flux suppression is well-modeled using He CIA.

The carbon abundances derived here for DQp stars appear to be smooth extension of the effective temperature sequence observed in the warmer DQ stars, indicating that the sudden disappearance of DQ stars near $T_{\rm eff}\approx6000$\,K is due to their smooth conversion to DQp spectral morphology (or rather the appearance of pressure distorted C$_2$ Swan bands). However, the carbon content of these stars depends on the amount of trace hydrogen, constrained here using CH upper limits for most stars.  However, ultraviolet data are best for highest accuracy, as done for 0038$-$226.

The consistent fits of the spectra by hydrogen-poor atmosphere models results in significantly higher densities than found in previous studies. In the coolest stars modeled here, with $T{\rm eff}<5000$\,K, these densities approach and may exceed 2\,g\,cm$^{-3}$.  Future modeling must account for the effects of high density on ionization equilibrium and CIA in these extreme stellar atmospheres.

\section*{Acknowledgments}
The authors are grateful to the anonymous reviewer for many suggestions that improved the manuscript, and in particular the suggestion to constrain trace atmospheric hydrogen content using the presence or absence of spectroscopic CH features.  The authors thank Steven Parsons for his painstaking reduction of the X-shooter data at a time when no pipeline existed, Richard Freedman for providing C$_2$ opacities, Patrick Dufour for sharing the details of his DQ white dwarf atmosphere models, and Stefano Bagnulo and John Landstreet for providing information on magnetic white dwarfs. The authors acknowledge computational resources provided by JARA-HPC partition project cjiek61. The authors also acknowledge the award of telescope time for ESO programs 097.D-0064 and 098.D-0059, and Gemini program GN-2008B-Q-110.

\section*{Author contributions}
JF and PMK contributed to the project equally.  PMK led all atmospheric modeling and spectral analysis, the interpretation of the results, and wrote the first draft.  JF is PI of the X-shooter data, performed the telluric corrections, led the writing and organization of the paper, and contributed some interpretation. SKL is PI of the NIRI data sets and performed all the necessary observations and data reduction.  HA \& JH collected and analyzed supplementary photometric measurements to support the analysis of the highly magnetic stars.  JPS contributed the ancillary, infrared-faint DC target 2008$-$600.

\section*{Data Availability}
The spectral data used in this paper are all available in the relevant observatory archives. The further post-processed and telluric-corrected spectra in the red optical and near-infrared are available on reasonable request to the authors.

\bibliographystyle{mnras}
\bibliography{references}

\end{document}